\documentclass[sigconf,authorversion,noacm,screen]{acmart}
\usepackage{fancyhdr}
\AtBeginDocument{\fancyhead[RO]{\sffamily\footnotesize Author's Version}}
\AtBeginDocument{\fancyhead[LE]{\sffamily\footnotesize Author's Version}}
\makeatletter
\def\blfootnote{\xdef\@thefnmark{}\@footnotetext}
\makeatother
\setcopyright{none}\renewcommand\footnotetextcopyrightpermission[1]{}

\copyrightyear{2024}
\acmYear{2024}
\setcopyright{rightsretained}
\acmConference[RAID 2024]{The 27th International Symposium on Research in Attacks, Intrusions and Defenses}{September 30-October 02, 2024}{Padua, Italy}
\acmBooktitle{The 27th International Symposium on Research in Attacks, Intrusions and Defenses (RAID 2024), September 30-October 02, 2024, Padua, Italy}
\acmPrice{}
\acmDOI{10.1145/3678890.3678907}
\acmISBN{979-8-4007-0959-3/24/09}

\usepackage{balance}
\usepackage[acronym,nohypertypes={acronym}]{glossaries}
\usepackage[utf8]{inputenc}
\usepackage[T1]{fontenc}
\usepackage{graphicx}
\usepackage{wrapfig}
\usepackage{rotating}
\usepackage{amsmath}
\usepackage{breakurl}
\usepackage{acmart-taps}
\usepackage{siunitx}
\usepackage{tikz}
\usepackage{marvosym} % not ACM-approved

\usepackage{xcolor}
\definecolor{nBlue}{RGB}{144, 167, 198}
\definecolor{nLightBlue}{RGB}{221, 229, 240}
\definecolor{nLightYellow}{RGB}{243, 238, 223}
\definecolor{nYellow}{RGB}{217, 198, 137}
\definecolor{nGreen}{RGB}{170,207,189}
\definecolor{nLightGreen}{RGB}{229,239,234}


\usetikzlibrary{arrows, calc, positioning, patterns}

\newacronym{AIO}{AIO}{asynchronous input/output}
\newacronym{ALU}{ALU}{arithmetic logic unit}
\newacronym{BPF}{BPF}{extended Berkeley Packet Filter}
\newacronym{CT}{CT}{constant time}
\newacronym{CET}{CET}{Control Flow Enforcement Technology}
\newacronym{CPU}{CPU}{central processing unit}
\newacronym{CRR}{CRR}{Connect/\allowbreak{}Request/\allowbreak{}Response}
\newacronym{DVFS}{DVFS}{dynamic voltage and frequency scaling}
\newacronym{IO}{IO}{input/\allowbreak{}output}
\newacronym{iTLB}{iTLB}{ instruction address translation lookaside buffer}
\newacronym{ISA}{ISA}{instruction set architecture}
\newacronym{JIT}{JIT}{just-in-time}
\newacronym{KPTI}{KPTI}{Kernel Page Table Isolation}
\newacronym{MPK}{MPK}{Memory Protection Key}
\newacronym{RR}{RR}{Request/Response}
\newacronym{RPS}{RPS}{Requests per Second}
\newacronym{STL}{STL}{Store To Load}
\newacronym{SCT}{SCT}{Speculative Constant-Time}
\newacronym{SNI}{SNI}{Speculative Non-Interference}
\newacronym{SMT}{SMT}{satisfiability modulo theories}
\newacronym{SLH}{SLH}{Speculative Load Hardening}
\newacronym{SSB}{SSB}{speculative store bypass}
\newacronym{SSBD}{SSBD}{Speculative Store Bypass Disable}
\newacronym{VM}{VM}{virtual machine}
\newacronym{vDSO}{vDSO}{a virtual Dynamically-linked Shared Object}
\newacronym{OS}{OS}{operating system}
\newacronym{OOB}{OOB}{out-of-bounds}
\newacronym{KASLR}{KASLR}{kernel address space layout randomization}
\newacronym{DPDK}{DPDK}{Data Plane Development Kit}
\newacronym{CNI}{CNI}{Kubernetes Container Networking Interface}
\newacronym{TCP}{TCP}{Transmission Control Protocol}
\newacronym{UDP}{UDP}{User Datagram Protocol}
\newacronym{SCTP}{SCTP}{Stream Control Transmission Protocol}
\newacronym{Wasm}{Wasm}{Web\-Assem\-bly}
\newacronym{PHT}{PHT}{Pattern History Table}
\newacronym{BTB}{BTB}{Branch Target Buffer}
\newacronym{RSB}{RSB}{Return Stack Buffer}

\date{\today}
\title{VeriFence: Lightweight and Precise Spectre Defenses for Untrusted Linux Kernel Extensions}
\hypersetup{
 pdfauthor={},
 pdftitle={VeriFence: Lightweight and Precise Spectre Defenses for Untrusted Linux Kernel Extensions},
 pdfkeywords={},
 pdfsubject={},
 pdfcreator={Emacs 29.3 (Org mode 9.6.15)}, 
 pdflang={English}}

\begin{document}

\author{Luis Gerhorst}
\orcid{0000-0002-3401-430X}
\affiliation{\institution{Friedrich-Alexander-Universität Erlangen-Nürnberg (FAU)}\country{Germany}}
\email{gerhorst@cs.fau.de}
\author{Henriette Herzog}
\orcid{0000-0002-0828-6862}
\affiliation{\institution{Ruhr-Universität Bochum (RUB)}\country{Germany}}
\email{henriette.herzog@rub.de}
\author{Peter Wägemann}
\orcid{0000-0002-3730-533X}
\affiliation{\institution{Friedrich-Alexander-Universität Erlangen-Nürnberg (FAU)}\country{Germany}}
\email{waegemann@cs.fau.de}
\author{Maximilian Ott}
\orcid{0009-0006-2374-5058}
\affiliation{\institution{Friedrich-Alexander-Universität Erlangen-Nürnberg (FAU)}\country{Germany}}
\email{ott@cs.fau.de}
\author{Rüdiger Kapitza}
\orcid{0000-0002-8116-7763}
\affiliation{\institution{Friedrich-Alexander-Universität Erlangen-Nürnberg (FAU)}\country{Germany}}
\email{ruediger.kapitza@fau.de}
\author{Timo Hönig}
\orcid{0000-0002-1818-0869}
\affiliation{\institution{Ruhr-Universität Bochum (RUB)}\country{Germany}}
\email{timo.hoenig@rub.de}

\begin{abstract}

  High-performance IO demands low-overhead communication between user- and
  kernel space. This demand can no longer be fulfilled by traditional system
  calls. %, which involve frequent transitions between user and kernel context.
  Linux's extended Berkeley Packet Filter (BPF) avoids user-/kernel transitions
  by just-in-time compiling user-provided bytecode and executing it in kernel
  mode with near-native speed. To still isolate BPF programs from the kernel,
  they are statically analyzed for memory- and type-safety, which imposes some
  restrictions but allows for good expressiveness and high performance. However,
  to mitigate the Spectre vulnerabilities disclosed in 2018, defenses which
  reject potentially-dangerous programs had to be deployed. We find that this
  affects
  \qty{31}{\percent} to
  \qty{54}{\percent} of
  programs in a dataset with
  844 real-world
  BPF programs from popular open-source projects. To solve this, users are
  forced to disable the defenses to continue using the programs, which puts the
  entire system at risk.

  To enable \textit{secure and expressive} untrusted Linux kernel extensions, we
  propose VeriFence, an enhancement to the kernel's Spectre defenses that
  reduces the number of BPF application programs rejected from
  \qty{54}{\percent} to
  zero.
  % TODO: add more key points here?
  We measure VeriFence's overhead for all mainstream performance-sensitive
  applications of BPF (i.e.,~event tracing, profiling, and packet processing)
  and find that it improves significantly upon the status-quo where affected BPF
  programs are either unusable or enable transient execution attacks on the kernel.

\end{abstract}


% https://dl.acm.org/doi/10.1145/3314221.3314590

% Constant-Time Foundations for the New Spectre Era:


\maketitle

\section{Introduction}\label{sec:intro}

\blfootnote{© 2024 Copyright held by the owner/author(s). This is the author's version of the work. It is posted here for your personal use. Not for redistribution. The definitive version was published in RAID'24.}

% \Aclp{OS} rely on system calls to allow the controlled communication of isolated
% processes with the kernel and other processes. Every system call includes a
% processor mode switch from the unprivileged user mode to the privileged kernel
% mode. Although processor mode switches are the essential isolation mechanism to
% guarantee the system’s integrity, they induce direct and indirect performance
% costs as they invalidate parts of the processor
% state \cite{gerhorst_anycall_2021}. In recent years, high-performance
% network \cite{noauthor_dpdk_2024} and storage hardware \cite{yang_spdk_2017}
% has made the user/kernel transition overhead the bottleneck for \acs{IO}-heavy
% applications. To make matters worse, security vulnerabilities in modern
% processors \cite{noauthor_refined_2022}, such as
% Meltdown \cite{lipp_security_2018}, have prompted kernel mitigations that
% further increase the transition overhead.

% To overcome this performance barrier, modern operating systems implement
% multiple alternatives to traditional blocking system calls. The three main
% approaches are \emph{asynchronous \ac{IO}} (\texttt{io\_uring} \cite{axboe_io_uring_2019}),
% \emph{kernel-bypass} (DPDK \cite{noauthor_dpdk_2024}), and \emph{safe kernel extensions}
% (BPF \cite{hoiland-jorgensen_express_2018}). This work focuses on safe kernel
% extensions because they offer lower latency than asynchronous
% \ac{IO} \cite{zhong_xrp_2022} and, unlike kernel-bypass, integrate well with
% the existing \ac{OS} networking stack \cite{hoiland-jorgensen_express_2018}.

Safe kernel extensions as implemented by Linux \gls{BPF} offer very low-overhead
interaction between user and kernel space. Their applications include network
\gls{IO} \cite{hoiland-jorgensen_express_2018,osinski_p4c-ubpf_2020,vieira_fast_2020},
memory optimization \cite{lian_ebpf-based_2022}, threat
detection \cite{kim_triaging_2022}, isolation \cite{jia_programmable_2023},
tracing \cite{yang_redis_2022}, scheduling \cite{heo_patchset_2023}, and
storage \cite{zarkadas_bpf-_2023}. While most applications today require root
privileges (\emph{privileged \gls{BPF}}), \gls{BPF} also allows unprivileged user processes
to load safety-checked bytecode into the kernel (\emph{unprivileged \gls{BPF}}), which
executes at near-native speed. Users typically develop \gls{BPF} programs in a
high-level programming language (e.g. C or Rust), which is compiled into
\gls{BPF} bytecode. This bytecode is verified in regard to its safety before
being just-in-time compiled. Invoking \gls{BPF} programs in the kernel and
calling kernel functions from within \gls{BPF} is much faster than the respective
switch to/from user context \cite{gerhorst_anycall_2021}. However, this
performance advantage comes at a cost: Executing untrusted code in the kernel
address space requires precise static analysis of the untrusted program to
maintain isolation. While some advanced safety-checks designed to prevent
transient execution attacks can be omitted for privileged \gls{BPF}, this is not
possible for unprivileged \gls{BPF}. Ensuring that these advanced defenses are
functional and low-overhead is therefore mandatory in order to enable the
various potential use cases for unprivileged \gls{BPF} (e.g., network traffic
filters~\cite{noauthor_unpriv-bpf_2024},
io\_uring~\cite{begunkov_io_uring_bpf_2021},
Seccomp~\cite{jia_programmable_2023}, and
others~\cite{edge:progress:2022,noauthor:hidbpf:2023}).

While Meltdown and similar domain-bypass transient execution attacks (e.g.,
where the hardware crosses address-space
boundaries \cite{noauthor_refined_2022}) can be efficiently prevented in
hardware, the cross/in-domain transient execution attacks discovered in
2018 \cite{kocher_spectre_2019} still require software defenses on all
high-performance \glspl{CPU} \cite{noauthor_managed_2018,mcilroy_spectre_2019} to date.
Two cross/in-domain transient execution attacks are a particular challenge for
\gls{BPF}: Spectre-PHT exploits the \acrlong{PHT}'s potential for branch
misprediction, and Spectre-STL exploits \acrlong{SSB} (via the \gls{CPU}'s \acrlong{STL}
buffer). Branches and stores are common in \gls{BPF} bytecode, but malicious
programs can use these very instructions to break out of their sandbox using
transient execution. For example, following a mispredicted branch, the program
can perform an unchecked \gls{OOB} access and leak the resulting kernel secret
using a covert channel. Preventing these attacks entirely in hardware requires
far-reaching changes to processor designs with significant performance
impacts \cite{xiong_survery_2021}
(e.g., \qtyrange{9}{15}{\percent} for \cite{yu_stt_2019}), that high-performance
\gls{CPU} vendors to date have not implemented \cite{amd:software:2023,intel:affected:2024,arm:speculative:2024}.

However, developing practical Spectre-resilient sandboxes is an open research
problem. To the best of our knowledge, \gls{BPF} is the only widely deployed
sandbox that attempts to implement full software-defenses against Spectre-PHT,
Spectre-BTB, and Spectre-STL. All other widely deployed sandboxes such as
Java \cite{naseredini_systematic_2021},
\gls{Wasm} \cite{dejaeghere_comparing_2023,wasmtime_security_2024}, and
JavaScript \cite{reis_site_2019,mcilroy_spectre_2019} instead rely on process
isolation. This is the approach recommended by
Intel \cite{noauthor_managed_2018} and V8
developers \cite{bynens_untrusted_2018}, but it is not applicable to \gls{BPF}.

\begin{figure}
\centering
\includegraphics[width=\columnwidth]{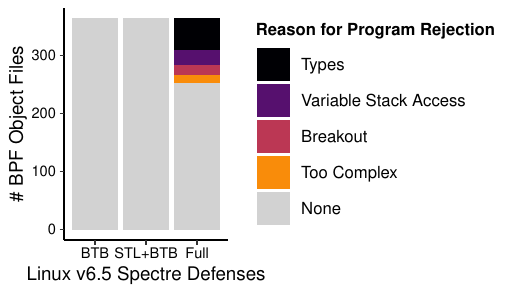}
\caption{\label{fig:orgbc73c4a}Reasons for program rejection in Linux when
  enabling defenses against transient execution attacks for
  364
  real-world \gls{BPF} test, example, and application object files. While the
  Spectre-\emph{BTB} and -\emph{STL} defenses do not trigger program rejections,
  enabling Spectre-PHT defenses (\emph{Full}, i.e., PHT+STL+BTB) prevents
  \qty{31}{\percent}
  of all objects (and even
  \qty{54}{\percent}
  of application objects) from being used.}
\end{figure}

To implement the sandbox, the kernel statically analyzes the control and data
flow of the \gls{BPF} programs before allowing them to execute in its address
space. Traditionally, the kernel only considered \gls{BPF} program paths that
could actually execute architecturally during this static analysis. In response
to the Spectre-PHT and Spectre-STL vulnerabilities disclosed in 2018, kernel
developers have extended the \gls{BPF} verifier \cite{ebpf_verifier} to prevent
leaks from transiently-executed \gls{BPF} program paths. This includes (a)
inserting instructions to make speculation safe (e.g., index
masking) \cite{starovoitov_bpf_2018,borkmann_801c60_2021,matei_01f810_2021},
(b) inserting instructions to prevent speculation (e.g., x86-64
\texttt{lfence}) \cite{borkmann_2039f2_2021}, or (c) statically analyzing the behavior
on speculative code paths to ensure they are safe, in turn rejecting the program
entirely if it exhibits any unsafe behavior \cite{borkmann_918367_2021}. The
latter is required to prevent Spectre-PHT, but it severely limits \gls{BPF}
because it prevents the user from loading their extension into the kernel
altogether.

To analyze the Spectre-PHT defenses' impact on \gls{BPF}, we collect
364 \gls{BPF} object files
(each containing one or multiple \gls{BPF} programs) from open-source projects and
enable defenses. The results, shown in Figure \ref{fig:orgbc73c4a}, confirm that
the number of rejections is significant and are further discussed in
Section \ref{sec:org26d664a}. Restructuring the program to not exhibit unsafe
transient behavior is tedious, as users are usually not familiar with transient
execution vulnerabilities and only indirectly control the compiler-generated
bytecode in their high-level sources.

Solving this problem of pessimistically rejected unprivileged programs would
make numerous new applications of \gls{BPF} practical. This is because
everything-but-trivial programs (approximately 10s of lines of C code) are
currently easily rejected, making development extremely tedious. It is our main
motivation to allow for more powerful and easier-to-develop unprivileged
programs and thereby enable the numerous future use-cases for unprivileged \gls{BPF}
that have been proposed by upstream kernel developers and academia:
  \begin{itemize}
    \item \textbf{Network Traffic Filters:} In upstream Linux, unprivileged \gls{BPF}
          is already allowed to filter network traffic
          \cite{noauthor_unpriv-bpf_2024}. Resolving the rejection issue
          will allow for more precise and easier-to-develop filters. These are
          extremely useful as demonstrated by the high number of projects that
          already utilize such filters
          \cite{hoiland-jorgensen_express_2018,osinski_p4c-ubpf_2020,vieira_fast_2020}
          but currently require root privleges.
    \item \textbf{io\_uring:} This will enable processes to utilize
          even better-performing asynchronous I/O due to a reduced
          number of switches to userspace
          \cite{begunkov_io_uring_bpf_2021}.
    \item \textbf{Seccomp:} This will allow unprivileged users to
          start processes that are restricted in regard to system calls and
          system-call parameters. In comparison to the existing solutions, it
          will allow for more precise filtering and thereby reduce the system's
          attack surface \cite{jia_programmable_2023}.
  \end{itemize}
  Aside from these three applications, there are various others
  \cite{edge:progress:2022,noauthor:hidbpf:2023} as unprivileged, safe kernel
  extensions are an extremely generic tool. All these applications of
  unprivileged \gls{BPF} can only become practical if the issues (i.e., program
  rejections) induced by the Spectre defenses are resolved.

    To solve this problem of pessimistically rejected unprivileged \gls{BPF} programs,
    our work extends the upstream \gls{BPF} verifier \cite{ebpf_verifier} by
    implementing VeriFence, an improved defense approach for Spectre. The core
    benefit of Veri\-Fence is that it reduces the number of programs that cannot
    be automatically mitigated and are, thus, rejected by the state-of-the-art
    \gls{BPF} verifier.

Our main contributions are fivefold:
\begin{itemize}
  \item \textbf{Domain Analysis}: To inform our design decisions, we statically
        analyze 844
        real-world \gls{BPF} programs from six popular software projects
        regarding their code size and complexity. This domain analysis gives us
        a detailed picture of the \gls{BPF} program landscape.
  \item \textbf{Security Notions:} Real-world Spectre defenses
        lack behind the theoretical foundations \cite{kim_ileakage_2023}.
        We discuss \gls{BPF}'s security in the light of
        transient execution attacks using established speculative security
        properties from the literature.
  \item \textbf{Design:} We design the VeriFence
        defense which optimistically attempts to verify all speculative
        execution paths and only falls back to speculation barriers when unsafe
        behavior is detected.
        % to eliminate the most common causes of program rejection using precise
        % defenses with a minimum expected performance impact.
\item \textbf{Evaluation}: We evaluate the performance impact of Veri\-Fence on the
  three most common use cases for \gls{BPF}: event tracing, continuous profiling,
  and network load balancing. We find that Veri\-Fence is lightweight, as it does not
  increase \gls{BPF}'s invocation latency.
\item \textbf{Implementation}: We publish Veri\-Fence for the v6.5 Linux kernel. Our
  patches soundly combine speculation barriers and static analysis to not increase the
  kernel's attack surface.
\end{itemize}

Besides our five main contributions, we discover multiple Linux kernel bugs
during our analyses. Based on our findings, we contribute proof-of-concept
exploits \cite{anonymous_exploit} and
fixes \cite{gerhorst_bpf_2023,gerhorst_082cdc_2023} that have been
accepted into upstream Linux.

\section{Background}
\label{sec:org03a1a6c}

This section discusses the transient execution vulnerabilities relevant to
\gls{BPF} and presents the respective defenses in Linux v6.5. We focus on
cross/in-domain transient execution attacks \cite{noauthor_refined_2022} but
not on domain-bypass transient execution attacks (including Meltdown /
Spectre v3 \cite{lipp_security_2018}, MDS \cite{schwarz:zombieload:2019},
L1TF \cite{bulck_foreshadow_2018}) because they can be efficiently addressed in
hardware and are not specific to \gls{BPF}. We focus on the \textbf{Spectre-PHT (v1),
-STL (v4), and -BTB (v2)} attacks because these are the longest-standing
vulnerabilities to which new processors are still
vulnerable \cite{amd:software:2023,intel:affected:2024,arm:speculative:2024}.

\subsection{Transient Execution Vulnerabilities}
\label{sec:org495a82b}

While regular (architectural) timing side-channel attacks can happen only when the
program computes explicitly on sensitive data (e.g., cryptographic keys, or also
kernel pointers with \acrlong{KASLR}, \acrshort{KASLR}), timing side-channel attacks based on
transient execution can also target victim code that does not explicitly work
with sensitive data. They become possible whenever a victim program encodes
secrets into side channels (e.g., the cache) during
transient execution.

\subsubsection{Speculation Triggers}
\label{sec:org7d8690e}

Transient execution \gls{CPU} vulnerabilities are commonly grouped by the
microarchitectural component that speculates. For \gls{BPF}, Spectre-PHT, -STL,
and -BTB are of particular
relevance \cite{canella_systematic_2019,cauligi_sok_2022}.
Spectre-PHT \cite{kocher_spectre_2019} includes all transient execution
vulnerabilities based on conditional branches as they use the \gls{PHT} for
target prediction. Spectre-STL exploits the fact that stores may not always
become visible to subsequent loads via the \gls{STL} buffer. While indirect
branches (which enable Spectre-BTB attacks \cite{kocher_spectre_2019}) also
occur in \gls{BPF}, they are not as common and the program cannot use them
directly. Because of this, the \emph{retpoline}-based defense against Spectre-BTB
is by default always enabled for privileged and unprivileged \gls{BPF} programs,
as it is also the case for the rest of the
kernel \cite{borkmann_bpf_2021,krysiuk_bpf_2022}. We therefore stick to the
default and always keep them enabled.

\subsubsection{Side Channels}
\label{sec:orga709350}

Transient execution attacks can use a variety of shared hardware components to
communicate sensitive data to attackers. Notably, this does not only include
caches but also microarchitectural execution \emph{ports} with simultaneous
multithreading \cite{bhattacharyya_smotherspectre_2019}. In this work, we
assume the established \textbf{\gls{CT} leakage model} to capture all these incidental
channels \cite{cauligi_sok_2022} by assuming that all accessed addresses are
potentially leaked. This includes both instruction- and data addresses, but not
the respective values at these addresses. From this, it follows that one must
not branch based on secret data nor use secrets as memory offsets.

\subsubsection{Unsafe Information-Flow}
\label{sec:org429fffd}

The CPU may even load additional sensitive data (erroneously) during speculative
execution. For example, \emph{speculative type
confusion} \cite{kirzner_analysis_2021} can cause the CPU to load sensitive
data from an attacker-controller address and subsequently leak the data through
a covert channel. To prevent any such unsafe information flow,
\gls{SCT} \cite{cauligi_constant-time_2020} and
\gls{SNI} \cite{guarnieri_spectector_2020} are notable formal notions that
capture the information-flow properties of the victim program that enable
transient execution attacks \cite{cauligi_sok_2022}. They are commonly referred
to as \emph{speculative security properties}.

\gls{SCT} extends the constant-time programming paradigm, which modern
cryptographic programs commonly use, to prevent Spectre gadgets by ensuring that the code executed
speculatively does not leak sensitive
data \cite{cauligi_sok_2022,cauligi_constant-time_2020}. A program that
satisfies \gls{SCT} must only operate on sensitive data using \gls{CT} processor
instructions \cite{noauthor_data_2023,noauthor_data_2023-1,noauthor_dit_2021}.
\gls{SCT} can be efficiently enforced using type systems that model the sensitive
data and its permitted
operations \cite{shivakumar_typing_2023,vassena_automatically_2021}.

\gls{SNI} formalizes the intuition that speculation can only leak data, which the
program already leaks during architectural
execution \cite{cauligi_sok_2022,guarnieri_spectector_2020,fabian_automatic_2022,patrignani_exorcising_2021}.
For example, if some scalar is already leaked into a side channel
architecturally, there is no point in protecting it from speculative leakage. To
enforce \gls{SNI} for a sandboxed program, \cite{guarnieri_spectector_2020} uses
symbolic execution and a \acrfull{SMT} solver. Like \gls{SCT}, \gls{SNI} not only
protects the sandbox runtime (e.g., the kernel) but also the sandboxed program itself
(e.g., a \gls{BPF} program) from Spectre attacks (i.e., they prevent \emph{poisoning
attacks}) \cite{cauligi_turning_2022}.

While enforcing either \gls{SCT} or \gls{SNI} prevents Spectre gadgets, \gls{SCT}
suffers from false positives while \gls{SNI} cannot be efficiently enforced for
arbitrary programs. Applying them to \gls{BPF} is therefore not necessarily
useful. Neither is applying them trivial as the verifier was not developed with
\gls{SCT} or \gls{SNI} in mind. However, in Section \ref{sec:orgfed59bc}, we
retrospectively analyze whether the \gls{BPF} verifier enforces \gls{SCT} or
\gls{SNI} for any of \gls{BPF}'s data types.

\subsection{Spectre Defenses of Linux \protect\gls{BPF}}
\label{sec:orgb797e88}
In this section, we present the first thorough analysis of \gls{BPF}'s Spectre
defenses. We build upon previous talks and reports on the
topic \cite{krysiuk_bpf_2022,borkmann_bpf_2021,schluter_security_2021} but
add more details and, in the following section, discuss whether \gls{BPF}
enforces any speculative security properties. We enable future work as
\gls{BPF}'s defenses to date are undocumented \cite{ebpf_verifier}. Also, the
number of contributors is
low \cite{borkmann_2039f2_2021,starovoitov_bpf_2018,matei_01f810_2021,borkmann_918367_2021}
and both our
work \cite{gerhorst_bpf_2023,anonymous_patch,anonymous_exploit,gerhorst_082cdc_2023}
and others \cite{borkmann_7fedb6_2021,borkmann_801c60_2021} have repeatedly
discovered bugs and misleading code.

\subsubsection{Whether to Enable Defenses for Privileged \protect\gls{BPF}}\label{sec:priv-bpf}

% Not:
  % \item
  %       For programs from unprivileged users, the kernel must verify both
  %       speculative and architectural security. For instance, the verifier must
  %       prevent unprivileged \gls{BPF} programs from printing pointer values in
  %       architectural execution, as well as prevent them from leaking pointers
  %       into covert channels during speculative execution.

While the need for applying Spectre defenses to unprivileged \gls{BPF} is clear, this % \enlargethispage{-25pt}
section presents the arguments for deciding whether to also enable defenses for
privileged users. As of today, privileged \gls{BPF} users are only restricted
regarding \gls{CPU} time (bounded loops) and memory usage (memory/type-safety).
\gls{BPF}'s Spectre-PHT and Spectre-STL defenses however are kept disabled by
default as they may lead to program rejections and are expected to have
performance overheads. With small programs, this is acceptable because
privileged users can be trusted to not insert Spectre gadgets into the kernel on
purpose.

With a rising number of privileged \gls{BPF} programs being used in production, there
is also a rising risk that gadgets from privileged programs are injected into
the kernel \emph{by mistake}. The kernel assumes that privileged users manually
check their programs for Spectre gadgets to prevent attacks on the \gls{BPF} program
and on the whole kernel. We argue this assumption is unrealistic, as most
developers are not familiar with transient execution attacks. Further, when
writing \gls{BPF} programs in high-level languages such as C, Rust, or Python, gadgets
are not directly visible and may only occur because of compiler optimizations.
Also, \gls{BPF} programs are frequently generated
ad-hoc \cite{xu_bpftrace_2024} and thus never undergo code review for
Spectre gadgets.

Figure~\ref{fig_stc} shows one example for a program vulnerable to Spectre-PHT
that may be loaded by an admin unfamiliar with transient execution
vulnerabilities. This effectively places an arbitrary-read gadget in the kernel
that can be used by anyone (even unprivileged or remote
users \cite{schwarz_netspectre_2018}) who controls the invocation of the \gls{BPF}
program, its input parameter \texttt{scalar}, and the covert channel used
(\texttt{covert\_channel[value]}).

\begin{figure}
\begin{verbatim}
I: // reg = is_ptr ? public_ptr : scalar;
A: if (!is_ptr) goto C; // mispredict
B: value = *reg;
   covert_channel[value];
C: exit();
\end{verbatim}
\captionof{figure}{Example from a \gls{BPF} program that contains a speculative type confusion gadget and is rejected by the Linux v6.5 \gls{BPF} verifier without Veri\-Fence.\label{fig_stc}}
\end{figure}

  To date, there exists no reliable static analysis tool that can find
    actually exploitable Spectre gadgets without false positives. Further,
    searching for these gadgets manually is extremely time-consuming. It is
    therefore an open question to which extent exploitable Spectre gadgets exist
    in real-world privileged \gls{BPF} programs. Our work allows cautious users to
    still eliminate this potential attack vector by enabling full Spectre
    defenses for privileged \gls{BPF} without risking program rejections.

\subsubsection{Attacker Model}\label{sec:orgd56e46a}

We consider both unprivileged attackers that can load arbitrary unprivileged
\gls{BPF} programs into the kernel, and attackers that can coerce a privileged
user into loading architecturally-safe privileged \gls{BPF} programs into the
kernel on their behalf. As of Linux v6.5, the kernel can only effectively
protect itself from the former.

We include the second type of attacker because many privileged programs are not
sufficiently reviewed for Spectre gadgets or generated
ad-hoc \cite{xu_bpftrace_2024}. They may therefore unintentionally contain
exploitable Spectre gadgets as we have discussed in the previous section.

\subsubsection{Architectural Safety}
\label{sec:orgb114e00}
To understand the existing Spectre defenses, which ensure speculative safety, we
first briefly present the design of the verifier and its mechanisms to ensure
architectural safety. The main goal of the verifier is to limit the damage
malicious or buggy \gls{BPF} programs can cause. For this, it verifies a bounded
execution time and memory safety but not functional correctness. Bounded
execution time is mainly useful for allowing \gls{BPF} use in interrupt contexts,
while memory safety prevents memory leaks and program bugs that would easily
enable kernel exploits.

To restrict data flow into the \gls{BPF} program's registers and stack, the
verifier enumerates all possible paths through the \gls{BPF} program and
simulates each path's execution. To verify memory and type safety, the kernel
analyzes the types (mainly different classes of pointers and
scalars \cite{starovoitov_bpftypes_2023}) and the value ranges of the scalars
(for which it uses \emph{tristate numbers} \cite{vishwanathan_sound_2022}). This
allows the verifier to ensure that all scalars used as pointer-offsets only
point to locations owned-by or borrowed-to\footnote{For example, this includes
pointers from the \texttt{bpf\_ringbuf\_reserve()} helper which must be either submitted
or discarded subsequently \cite{noauthor_bpf2_2023}.} the \gls{BPF} program. For
example, this prevents \gls{OOB} accesses to the \gls{BPF} stack and to network
packets (the \gls{BPF} program can manipulate packets directly using its context
pointer \cite{anonymous_patch}). Further, accessing uninitialized stack slots and
registers is also prohibited. In summary, the \gls{BPF} program can only access
memory locations (i.e., load their value into registers) to which the kernel
grants explicit access.

The kernel not only restricts how data flows into \gls{BPF} program registers and
stack but also limits how the program uses the data in these locations. For
this, the verifier distinguishes between pointers and scalars. This allows the
verifier to limit how the \gls{BPF} program processes data and ensure that only
safe operations are executed. For example, programs can only dereference
pointers (at valid offsets) and not cast them into scalars. Further, they can
only use scalars in arbitrary \gls{ALU}~operations and conditional branching
based on their value \cite{anonymous_patch}. Therefore, the \gls{BPF} program
can only use kernel pointers in \gls{CT} operations (except for dereferencing
them), while scalars are only restricted so that the program cannot cast them
into pointers.

\subsubsection{Design Goals}
\label{sec:org23a0d71}

By extending the design used to ensure architectural safety, the kernel's Spectre
defenses for \gls{BPF} also enforce that the same restrictions apply in speculative
execution as well. The goal in Linux v6.5 is only to protect the kernel from
\gls{BPF} programs and associated user applications. The \gls{BPF} program itself
is not protected against other user applications that might exploit Spectre
gadgets to retrieve scalars (including cryptographic keys) the program
processes. Further, the existing defenses against Spectre-PHT and Spectre-STL
try to be transparent to users as much as possible, as the kernel applies them
to the bytecode during verification time. The kernel does not require the user
or source compiler to insert speculation barriers into the program. However, the
defenses fail to be entirely transparent, as they can impact performance and
prevent programs from passing verification.

\subsubsection{Spectre-STL}\label{sec:spectre-stl}

To defend against Spectre-STL (v4), the kernel inserts speculation barriers
after \emph{critical} stores to the \gls{BPF} stack \cite{borkmann_2039f2_2021}. A
store is \emph{critical} if speculatively bypassing it would make otherwise
inaccessible data (or operations upon data) available to the program. For
example, initializing a stack slot or overwriting a scalar value with a pointer
are critical stores as the kernel prohibits reading uninitialized stack slots
and dereferencing scalars. The verifier can only skip the insertion of a barrier
if a scalar is overwritten with another scalar \cite{gerhorst_bpf_2023}.
Confusing one scalar value with another scalar cannot lead to \gls{OOB} memory
accesses since the verifier enforces pointer limits using branchless logic
(e.g., masking). This is further discussed in the following section, as it is
also required to defend against Spectre-PHT.

This defense is expected to have an impact on \gls{BPF}'s performance as the verifier
inserts speculation barriers that reduce instruction-level parallelism (even if
there is no misprediction). However, this impact has not been measured
exhaustively for real-world applications, and we therefore include it in our
evaluation.

\subsubsection{Spectre-PHT}
\label{sec:orga0f2aca}
While Spectre-STL effectively causes the \gls{CPU} to speculatively bypass a
store, Spectre-PHT (v1) causes the \gls{CPU} to bypass evaluation
of a branch condition. Therefore, conditional branches can no longer be reliably
used to ensure memory and type safety. To defend against this, the verifier
prevents \gls{OOB} accesses using branchless
logic \cite{starovoitov_bpf_2018,borkmann_801c60_2021,matei_01f810_2021} and
type confusion by verifying architecturally-impossible speculative execution
paths \cite{borkmann_918367_2021}.

\emph{Branchless Bounds Enforcement.} This includes simple masking but also more
complex instruction sequences when required. First, for arrays (i.e., \gls{BPF}
maps) the kernel can simply ceil the size to a power of two and apply the
respective index-mask before the access \cite{starovoitov_bpf_2018}. Second,
for pointers, the kernel deducts the bounds from the conditional branches that
lead to the pointer dereference. Then the kernel also enforces them directly
before the access using a special sequence of \gls{ALU} operations \cite{borkmann_801c60_2021} (additionally, the verifier has to check
a remaining corner case using verification of an architecturally impossible
speculative code path). Third, for the stack, the kernel enforces that all
offsets are constant to simplify the implementation \cite{matei_01f810_2021}.
Finally, there is one exception where the kernel allows the program to access
\gls{OOB} memory speculatively for practical
reasons \cite{anonymous_maintainer_re_2023}. We have verified that the specific
memory layout in this case does not expose any
secrets \cite{anonymous_exploit}.

\emph{Verification of Speculative Execution Paths.} While branchless bounds
enforcement prevents the \gls{BPF} program from accessing forbidden data, it
cannot prevent the program from using kernel pointers or scalars in an unsafe
manner (e.g., dereferencing a scalar). To ensure speculative type safety, the
kernel also simulates and checks the execution paths that include mispredicted
branches \cite{borkmann_918367_2021}. The verifier only allows transient
behavior that is also permitted architecturally. Even though this is a sound
approach, it suffers from false positives because not every architecturally
unsafe operation enables a Spectre attack \cite{guarnieri_spectector_2020} (and
the architectural verification logic itself also already suffers from false
positives). Using separate simulation and verification logic for the transient
domain could resolve this, but it would increase the verifier's attack surface
even further.

In summary, branchless bounds enforcement and verification of mispredicted
branches enable reliable defense against Spectre-PHT.

\section{Security Analysis}
\label{sec:orgfed59bc}
In this section, we analyze the foundations of \gls{BPF}'s Spectre defenses.
We do this both from a hardware and a formal perspective. Regarding the
hardware, we summarize the required instruction-set properties for the defenses
(i.e., the hardware-software contract). Further, we retrospectively
analyze the speculative security properties the verifier enforces.

\subsection{Hardware-Software Contract}
\label{sec:orga18b8c0}

To terminate speculative execution after an \gls{STL} misprediction, the verifier
relies on speculation barriers. For this, it uses the undocumented \texttt{nospec}
\gls{BPF} bytecode instruction that is not available to user space. Notably, on
x86-64, the \gls{JIT} implements these using the \texttt{lfence} instruction, which is
in line with Intel's recommendation for Spectre-STL and
-PHT \cite{noauthor_speculative_2018,noauthor_managed_2018}. Independent
research has confirmed that \texttt{lfence} terminates
speculation \cite{mambretti_speculator_2019}, therefore, we deem them reliable.
On ARM64, the \gls{JIT} compiler lowers \texttt{nospec} to a no-op because the hardware
already defends against Spectre-STL in firmware \cite{noauthor_firmware_2018}.
To perform safe pointer arithmetic, the verifier further relies on \gls{ALU} instructions that have data-independent
timing \cite{noauthor_dit_2021,noauthor_data_2023}.

\subsection{Speculative Security Properties}\label{sec:spec-sec-prop}

The verifier enforces a mix of speculative security properties for \gls{BPF}
programs. Overall, the verifier attempts to prevent the \gls{BPF} program from
speculatively breaking out of its sandbox and is successful in that (to the best
of our knowledge), except for the exception discussed in
Section \ref{sec:orga0f2aca} \cite{anonymous_exploit}.
\emph{Speculative-breakout attacks} are therefore prevented. Regarding the
exception, we are also unable to construct an exploit that leaks
\emph{sensitive} kernel data \cite{anonymous_exploit}. However, while this in
summary protects the kernel, the verifier does not enforce
\gls{SNI} \cite{cauligi_sok_2022} as processed scalars, which are never leaked
architecturally, can still leak after a speculative scalar confusion due to
Spectre-STL. Therefore, there is no support for \gls{BPF} programs in protecting
cryptographic keys they process. While users cannot insert speculation barriers
manually in Linux v6.5, the kernel could offer basic support for this by
exposing the internal \gls{BPF} bytecode instruction to insert barriers
(\texttt{nospec}) to users. Regarding \gls{SCT}, the verifier enforces it only
for pointers, but with the exception that the program can dereference the
pointer itself. Overall, retrofitting a single formal speculative security
property to \gls{BPF} does not appear viable. However, we still find \gls{SNI} and
\gls{SCT} to be of high value as we discover multiple kernel bugs by analyzing
\gls{BPF}'s defenses with them in mind
\cite{gerhorst_bpf_2023,gerhorst_082cdc_2023,anonymous_patch}.

\section{Problem Statement}\label{sec:problem}

Out of the limitations we have identified throughout the previous sections, we
deem the Spectre-PHT defenses leading to program rejections to be the most
limiting to users: If the verifier finds a speculative execution path (following
a simulated misprediction) that performs prohibited operations, it rejects the
whole program \cite{borkmann_918367_2021}. This forces users to resort to
tedious restructuring of the source code or abandon \gls{BPF} completely (which
has a much higher performance impact than the speculation
barriers \cite{gerhorst_anycall_2021}). Frequently, this development overhead
motivates users to simply disable defenses
altogether \cite{anonymous_user_re_2023}, which enables \gls{BPF}-based exploits
as shown in the original Spectre paper \cite{kocher_spectre_2019}.

With Spectre-PHT identified as the most limiting problem, we focus on the
respective rejections in this work and leave solving the lack of support for
\gls{SNI} (Section~\ref{sec:spec-sec-prop}) and the performance
regressions due to Spectre-STL defenses (Section~\ref{sec:spectre-stl}) to
future work.

\section{Domain Analysis}
\label{sec:orgdd58b36}
As discussed in the previous section, Linux v6.5's Spectre-PHT defenses cause
\gls{BPF} program rejections when the verifier cannot prove some transient
execution is safe. In this case it prevents unprivileged users from using
\gls{BPF} altogether. At the same time, the
Spectre-STL defenses can negatively impact the execution time as they insert
speculation barriers. However, to-date it is
unclear to which extent these defenses are actually triggered in real-world
\gls{BPF} programs.

In this section, we present a domain analysis of the \gls{BPF} program landscape
to analyze how \gls{BPF} programs are affected by the kernel's Spectre defenses.
We collect 364 \gls{BPF}
object files containing
844 individual
BPF programs from six popular open-source projects and analyze the number of
objects rejected and speculation barriers inserted. To support future research,
we publish our tools for collecting and analyzing the
programs.\footnote{\anon[Reference omitted for double-blind
  review.]{\url{https://sys.cs.fau.de/verifence}}}

\subsection{Dataset}
\label{sec:orgb217fa3}

We include a diverse set of programs taken from the Linux kernel selftests and
BPF samples \cite{kroah-hartman_linux_2023}, libbpf
examples \cite{nakryiko_libbpflibbpf-bootstrap_2022},
Prevail \cite{gershuni_simple_2019}, BCC \cite{xu_bcc_2023}, the Cilium
\gls{CNI} \cite{stringer_ciliumcilium_2022}, the Parca Continuous
Profiler \cite{javier_honduvilla_coto_parca-devparca-agent_2023}, and the
Loxilb Network Loadbalancer \cite{packetcrunch_github_user_loxilb_2024}\footnote{Despite our best efforts, we are unable to use most of the \gls{BPF} objects from
the Prevail paper because they are not compatible with Linux v6.5. For the
Cilium project, only the \texttt{bpf\_sock} program is compatible with our
\texttt{bpftool}-based toolchain \cite{torvalds_bpftool_2022}.}.
We observe that programs from the Linux kernel selftests are often designed to
only test a specific kernel interface and are therefore very small. This motivates
us to further group the programs into \textbf{test or example} and \textbf{application}
programs
(171 programs
from
50 object
files). The application group only includes programs from the BCC, Loxilb,
Parca, and Cilium projects, as well as
2 specific
object files from the Linux selftests with programs adapted from real-world
applications.

\subsubsection{Applicability to Unprivileged \protect\gls{BPF}}\label{sec:apl-to-unpriv-bpf}

Unfortunately, because of their high risk for rejection by the verifier, few
real-world \gls{BPF} programs designed for unprivileged use exist today. We aim to
enable more real-world \gls{BPF} programs by reducing unnecessary program rejections.
To still ensure a reasonably-sized dataset for our evaluation, we modify the
kernel to support activation of Spectre-PHT and -STL defenses at runtime and
then also enable these defenses for existing privileged \gls{BPF} programs we collect
from the open source projects.

Still, our evaluation accurately represents the real-world unprivileged \gls{BPF}
programs that will be developed for our motivating use-cases from
Section~\ref{sec:intro} in the future. This is because, at a bytecode-level, the
program behaviors that lead to Spectre induced program rejections or
performance-regressions are the same for unprivileged and privileged \gls{BPF}, as
they primarily differ in the kernel interfaces (i.e.,~attachment point and \gls{BPF}
helper function calls) they interact with. The compiler toolchain, kernel
infrastructure, and high-level verification algorithm are exactly identical for
privileged and unprivileged \gls{BPF}. Programs loaded for privileged and unprivileged
use only differ in that the verifier ignores minor unsafe behaviors for
privileged programs (basically assuming they must be false positives as the user
is trusted) while triggering a rejection for programs from unprivileged users.

Upstream kernel developers have also used this technique to analyze the impact
of the Spectre-PHT defenses on pointer arithmetic \cite{borkmann_bpf_2019}.
We have validated the maintainer's reasoning here and see no technical
  reason why the conclusions drawn from our dataset should not apply to future
  unprivileged \gls{BPF} programs.

\subsubsection{Program Size}
\label{sec:orgd47e6ba}

To inform our analysis, we first measure the number of \gls{BPF} bytecode
instructions per program. We observe that the median number of instructions per
program is only
40 overall.
For the programs classified as applications, the median is
46 instructions
per program, while the arithmetic mean is
559 since
there are some programs that are close to the verifier's complexity limit of
\num{1e6} instructions (e.g., Parca's stack sampler). Overall, the low median
number of instructions is expected as most \gls{BPF} programs only implement a
fast path or policy decision in the kernel.

\subsection{Speculation Barriers}
\label{sec:org1713803}

\begin{figure}
\centering
\includegraphics[width=\columnwidth]{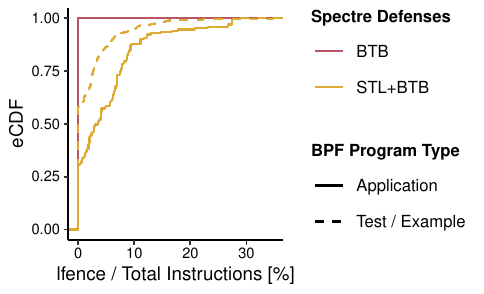}
\caption{Percentage of speculation barriers in 844 \gls{BPF} programs as a fraction of the total number of machine instructions with Spectre-STL defenses active.\label{fig_lfence}}
\end{figure}

We first analyze how many speculation barriers the verifier inserts to defend
against Spectre-STL. Figure~\ref{fig_lfence} shows the percentage of barriers per
\gls{BPF} program for all collected object files that are compatible with our
kernel version. Comparing the programs classified as test or example programs
with those from application projects, we find that application programs require
more barriers per instruction on average
(\qty{2.2}{\percent} instead
of
\qty{1.0}{\percent} overall).
This is likely because they are usually more complex and thus cannot work with
registers exclusively. Simply counting the number of speculation barriers, of
course, only indirectly relates to a real-world performance overhead, which
heavily depends on the exact location of the barrier (e.g., in a tight loop or
only at the beginning of the program during initialization). In
Section \ref{sec:orgf874708}, we will, therefore, analyze the performance overhead of the
Spectre-STL defenses in real-world execution-time benchmarks. Still, we expect
the overall overhead to be much lower compared to the performance overhead if
\gls{BPF} were not used at all \cite{gerhorst_anycall_2021}.

\subsection{Program Rejections}
\label{sec:org26d664a}
While the Spectre-STL defenses of Linux v6.5 only slow down the \gls{BPF}
program, the Spectre-PHT defenses can lead to the entire program being rejected.
This forces users to resort to tedious restructuring of the source code or
abandon \gls{BPF} completely. In this section, we analyze how many programs from
real-world projects the verifier actually rejects.

Figure \ref{fig:orgbc73c4a} from Section \ref{sec:intro} shows the number and
cause of program rejections with only Spectre-BTB defenses active (\emph{BTB}),
Spectre-STL and -BTB defenses active (\emph{STL+BTB}), and with additional
Spectre-PHT defenses active (\emph{Full}, i.e., PHT+STL+BTB). As expected, the
Spectre-STL defenses cause no rejections. However, with the
  Spectre-PHT defenses active, the verifier rejects
  \qty{31}{\percent}
  of the
  364
  compatible \gls{BPF} object files we have collected. Excluding test or example
programs, we even find that
\qty{54}{\percent} of
the remaining
50 application
object files are rejected (likely because they are usually more complex and,
therefore, more likely to contain any unsafe speculative behavior). We group the
causes for rejections into four categories:
\begin{itemize}
  \item \textbf{Types
        (\qty{15}{\percent}):}
        These are type errors on speculative program paths. For example,
        speculatively casting as scalar into a pointer and dereferencing it.
  \item \textbf{Variable Stack Access
        (\qty{7}{\percent}):}
        The program contains a speculative or architectural stack access for
        which the verifier cannot statically compute the offset. To simplify the
        Spectre-PHT defenses, kernel developers have disallowed variable-stack
        accesses \cite{matei_01f810_2021}.
  \item \textbf{Breakout
        (\qty{5}{\percent}):}
        Unsafe speculative execution where the \gls{BPF} program accesses
        locations (memory or uninitialized registers) owned by the kernel.
  \item \textbf{Too Complex
        (\qty{4}{\percent}):}
        Because the verifier has to check the additional speculative paths, some
        programs exceed the verifier's complexity limits (e.g., a maximum path
        length of \num{1e6} instructions).
\end{itemize}
In summary, the verifier rejects \gls{BPF} programs due to a diverse set of
unsafe behaviors that can lead to transient execution attacks on the kernel if
the program leaks the resulting secret into a side channel subsequently.

In Table~\ref{tab_load_problems_projects}, we further analyze the number of
rejections per software project. As expected, there is no notable difference in
the extent to which the rejections affect the different projects since most of
the bytecode-level properties that lead to unsafe speculation do not directly
map to high-level constructs in the source code. The only exceptions are
architectural variable-stack accesses \cite{matei_01f810_2021}, which users can
avoid by not allocating arrays on the stack. In summary, all but one of the
projects we have analyzed (that is Cilium, for which we only have one object file
compatible with our toolchain) are negatively affected by the Spectre-PHT
defenses.

\begin{table}
  \centering
  \caption{Number of \gls{BPF} object files rejected with the existing Linux v6.5 defenses for different software projects.}\label{tab_load_problems_projects}
  \begin{tabular}{ l r r r }
    \toprule
    \textbf{Project} & \textbf{\# Programs}  & \textbf{\# Files} & \textbf{\# Files Rejected} \\
    \midrule
    Linux Selftests&592&275&80 \\ 
BCC&133&39&19 \\ 
Linux Samples&71&32&5 \\ 
Loxilb&19&4&3 \\ 
Cilium&10&1&0 \\ 
libbpf Examples&10&7&1 \\ 
Parca&7&4&3 \\ 
Prevail&2&2&1 \\ 

    \bottomrule
  \end{tabular}
\end{table}

To conclude, we find that the Spectre-PHT defenses in Linux v6.5 are
significantly more limiting to users than the Spectre-STL defenses, showcasing
the significance of our work. The defenses cause
users to resort to disabling defenses altogether (e.g., \cite{anonymous_user_re_2023}), which opens the door to introducing dangerous
arbitrary-read gadgets into the kernel. In the following section, we present
Veri\-Fence, which defends against Spectre-PHT without rejecting the whole
\gls{BPF} programs.

\section{Design}
\label{sec:orgc9c71d3}
As outlined in the previous sections, the current \gls{BPF} verifier
pessimistically rejects programs that pose a potential security threat. While
this is preferable to leaving the system vulnerable to transient execution attacks, it also limits \gls{BPF}'s usability. These rejections seem unnecessary,
especially considering that effective defense mechanisms exist, which could be
included in the program instead.

In this section, we present the idea and design of \textbf{Veri\-Fence}, not only a
solution to the problem from Section~\ref{sec:problem}, but also a generic technique to build Spectre-resistant
software sandboxes based on verification. Veri\-Fence optimistically
attempts to verify all speculative execution paths and only falls back to
speculation barriers when unsafe behavior is detected. Importantly, we fully
reuse existing verification logic for this to balance verifier complexity (and
thereby potential for kernel bugs) with \gls{BPF} execution-time overheads.
Surprisingly, our evaluation shows that this approach results in low application
overheads because invocation-latency is more critical to \gls{BPF}'s performance
than code execution-time. While we only implement Veri\-Fence's approach for
Spectre-PHT, it can also be used to reduce the number of
speculation barriers required for Spectre-STL
defenses \cite{borkmann_2039f2_2021,gerhorst_bpf_2023}, and to Spectre-BTB
with Intel's \gls{CET} \cite{shanbhogue_cet_2019}. We first present our idea by
example and then discuss its security and performance.

\subsection{Fence or Verify}
\label{sec:orgce3f59d}

\begin{figure}
	\begin{tikzpicture}

	%%%% Background %%%%
	\draw[fill=nLightBlue, draw=nLightBlue] (0,0) rectangle ++(-55pt,-140pt)
		node[rectangle, align=center, pos=0.5, yshift=-60pt] {\textsf{\textbf{Path 1}}};
	\draw[fill=nLightYellow, draw=nLightYellow] (0,0) rectangle ++(55pt,-140pt)
		node[rectangle, align=center, pos=0.5, yshift=-60pt] {\textsf{\textbf{Path 2}}};

\node[draw, circle, fill=white] at (0,-0.4) (i) {\textsf{I}};

%	\draw[dashed] (i.south) -- ([yshift=-120pt]i.south);

	%%%% Path 1 %%%%
	\node[draw, fill=white, circle, below left=20pt and 10pt of i] (a1) {\textsf{A}};
	\node[draw, fill=white, circle, below=15pt of a1] (b1) {\textsf{B}};
	\node[draw, fill=white, circle, below=15pt of b1] (c1) {\textsf{C}};

	\draw[->] (i) -- (a1) node [midway, above, xshift=-20pt] {\footnotesize\texttt{reg=pub\_ptr}};
	\draw[->] (a1) -- (b1) node [midway, left] {\footnotesize\texttt{is\_ptr}};
	\draw[->] (b1) -- (c1);

	%%%% Path 2 %%%%
	\node[draw, fill=white, circle, below right=20pt and 10pt of i] (a2) {\textsf{A}};
	\node[draw, fill=white, circle, below=46.5pt of a2] (c2) {\textsf{C}};

	\draw[->] (i) -- (a2) node [midway, above, xshift=20pt] {\footnotesize\texttt{reg=scalar}};
	\draw[->] (a2) -- (c2) node [midway, right, yshift=16pt] {\footnotesize\texttt{$\neg$is\_ptr}};

\end{tikzpicture}
\hfill
\begin{tikzpicture}

	%%%% Background %%%%	
	\draw[preaction={fill=nLightBlue}, draw=nLightBlue, pattern=north west lines, pattern color=nBlue!80!white] (0,0) rectangle ++(-55pt,-140pt)
		node[rectangle, align=center, pos=0.5, yshift=-60pt] {\textsf{\textbf{Path 3}}};
	\draw[preaction={fill=nLightYellow}, draw=nLightYellow, pattern=north west lines, pattern color=nYellow!80!white] (0,0) rectangle ++(55pt,-140pt)
		node[rectangle, align=center, pos=0.5, yshift=-60pt] {\textsf{\textbf{Path 4}}};

	\node[draw, circle, fill=white] at (0,-0.4) (i) {\textsf{I}};

%	\draw[dashed] (i.south) -- ([yshift=-120pt]i.south);

	%%%% Path 3 %%%%
	\node[draw, fill=white, circle, below left=20pt and 10pt of i] (a1) {\textsf{A}};
	\node[draw, fill=white, circle, below=46.5pt of a1] (c1) {\textsf{C}};

	\draw[->] (i) -- (a1) node [midway, above, xshift=-20pt] {\footnotesize\texttt{reg=pub\_ptr}};
	\draw[->, dash dot] (a1) -- (c1) node [midway, left, yshift=16pt] {\footnotesize\texttt{is\_ptr}};

	%%%% Path 4 %%%%
	\node[draw, fill=white, circle, below right=20pt and 10pt of i] (a2) {\textsf{A}};
	\node[draw, fill=white, circle, below=15pt of a2] (b2) {\textsf{B}};
	\node[draw, fill=white, circle, below=15pt of b2] (c2) {\textsf{C}};
	
	\node[left=-1pt of b2] {\huge\Lightning}; 

	\draw[->] (i) -- (a2) node [midway, above, xshift=20pt] {\footnotesize\texttt{reg=scalar}};
	\draw[->, dash dot] (a2) -- (b2) node [midway, right] {\footnotesize\texttt{$\neg$is\_ptr}};
	\draw[->] (b2) -- (c2);

\end{tikzpicture}
  \caption{Execution paths through the \gls{BPF} program from Figure~\ref{fig_stc}. Paths marked with $\rightarrow$ denote architectural execution, while a dashed arrow indicates speculative execution. \begin{imageonly}\Lightning{}\end{imageonly}~indicates unsafe behavior the verifier must prevent.}
	\label{fig:04-paths}
\end{figure}
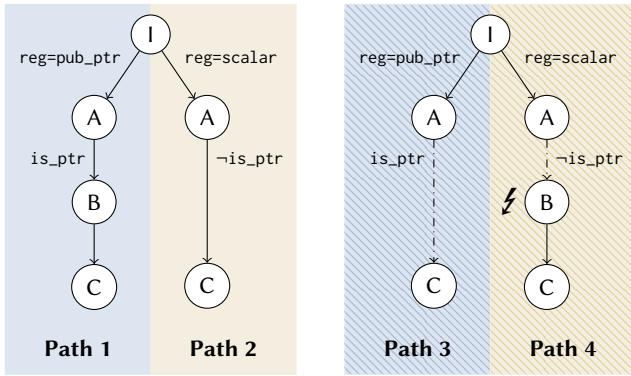

Based on our analysis of the \gls{BPF} program landscape, we have identified the
existing Spectre-PHT defenses to be the main real-world issue because they
prevent applications from using \gls{BPF} altogether. In this section, we analyze
a minimal example, shown in Figure~\ref{fig_stc}, of a \gls{BPF} program where
Veri\-Fence successfully prevents the critical transient execution
while the Linux v6.5 \gls{BPF} verifier rejects the entire
program \cite{borkmann_918367_2021}. The program executes either with \texttt{is\_ptr}
true or false, and we assume logic in block \texttt{I} implements \texttt{reg = is\_ptr ?
public\_ptr : scalar}. We exclude the corresponding branch in block \texttt{I} in
this analysis to give a concise example. Figure~\ref{fig:04-paths} illustrates all
execution paths through the \gls{BPF} program. Path 1 and Path 2 are
architecturally possible paths through the program. By simulating the execution
of each basic block, the verifier computes the invariants that will hold after
executing the block based on the inputs, the block's instructions, and the
condition of the final branch that ends the basic block (e.g., \texttt{reg} will be
equal to \texttt{public\_ptr} and \texttt{is\_ptr} must be true, when we go to block \texttt{B}
after \texttt{A}). We list these invariants on the edges connecting the basic
blocks in Figure~\ref{fig:04-paths}.

Only considering the architectural execution paths, the program could be
accepted because the verifier understands that block \texttt{B} will only
execute if \texttt{reg} contains a valid pointer (in Path 1). Thus, the program
does not exhibit any architecturally unsafe behavior and was therefore accepted
by the kernel before 2018. However, with the defenses active
(specifically \cite{borkmann_918367_2021}), the program is rejected as the
verifier also simulates branch mispredictions, which leads to the speculative
execution paths (Path 3 and 4) shown in Figure~\ref{fig:04-paths}. With Spectre
defenses, the verifier rightly finds the program to be unsafe because it
performs a forbidden pointer dereference on Path 4. Specifically, when
block \texttt{B} executes with a scalar in \texttt{reg}, the program breaks out of it
sandbox by accessing an arbitrary address (\begin{imageonly}\Lightning{}\end{imageonly} in Figure~\ref{fig:04-paths}).
Subsequently, the resulting kernel secret could be leaked into a covert channel
(denoted as \texttt{covert\_channel[value]} in the code example). In the Linux v6.5
kernel, this unsafe behavior causes the whole \gls{BPF} program to be rejected,
which either prevents unprivileged users from using the program altogether or
forces privileged users to disable the Spectre defenses.

Our solution, Veri\-Fence, solves this by dynamically falling back to inserting
a speculation barrier whenever the verifier detects unsafe program behavior
after a simulated misprediction. By utilizing speculation barriers instead of
program rejection to protect against unsafe program behavior, VeriFence
successfully addresses Section~\ref{sec:problem}. Further, whenever
  the verifier encounters a barrier while verifying a speculative path, it can
  stop verifying this path. Both of these modifications are based upon the
  insight that speculation barriers behave like program exits when the execution
  is transient. In the example from Figure~\ref{fig_stc}, Veri\-Fence triggers the
  insertion of a barrier at the beginning of block \texttt{B} before
  \texttt{reg} is dereferenced. This still allows the CPU to mispredict the
  branch but terminates the speculative execution path when block \texttt{B} is
  reached. In contrast, the speculative execution in Path 3 is not problematic
  because no unsafe operation is performed in block \texttt{C}, even though the
  CPU mispredicted the branch at the end of block \texttt{A}. Veri\-Fence
  rightly detects this and does not insert a speculation barrier at the
  beginning of block \texttt{C}.

\subsection{Security}
\label{sec:orgdf7806e}

Veri\-Fence only allows transient behavior that was already allowed
architecturally and prevents all other execution paths using speculation
barriers. We can therefore reduce its security to the security of the existing
\gls{BPF} verifier. Importantly, this allows Veri\-Fence to benefit from
formal verification efforts that improve the security of the verifier
\cite{vishwanathan_sound_2022,nelson_jitterbug_2020}, even if they do not explicitly take Spectre
into account. In summary, Veri\-Fence uses a reliable and easy-to-implement
method to prevent unsafe speculative behavior (and thereby the use of
secret-leaking side- and covert channels).

Our extension does not change the speculative security properties enforced by
the \gls{BPF} verifier, because we do not permit any new speculative behavior.
With the assumptions noted in Section \ref{sec:orgfed59bc}, our modified verifier
is therefore still \emph{secure}, assuming the \gls{CT} leakage model, an unprivileged
local attacker, and the secrecy policy identified
in~Section~\ref{sec:orgb114e00}.

While Veri\-Fence does not affect security against unprivileged attackers, it
improves security in the light of privileged users that only manually verify
architectural safety. With Veri\-Fence, these users no longer accidentally load
\gls{BPF} programs into the kernel that unexpectedly leak sensitive kernel or
user data. For example, loading the program from Figure~\ref{fig_stc} into the
kernel effectively places an arbitrary-read gadget in the kernel that can be
used by anyone (even unprivileged or remote
users \cite{schwarz_netspectre_2018}) who controls the invocation of the
\gls{BPF} program, its input parameter \texttt{scalar}, and the covert channel
used (\texttt{covert\_channel[value]}). With Veri\-Fence enabled for privileged
users (which becomes practical because of the improved expressiveness), no
arbitrary data will be read into \texttt{value} and therefore the kernel remains
secure.

In summary, Veri\-Fence soundly prevents unsafe transient execution and directly
benefits from work that discovers architectural verifier bugs. The main expected
downside of Veri\-Fence's approach, in comparison to a vulnerable system, is
the execution-time overhead at runtime due to the speculation barriers and the
added verification-time overhead. Both are the topic of the following section
and are further analyzed in the evaluation.

\subsection{Performance}
\label{sec:org1cd9b35}

In this section, we analyze Veri\-Fence's impact on performance. In particular,
Veri\-Fence has an impact on the execution time of the \gls{BPF} program and
on the verification time when loading a \gls{BPF} program.

Veri\-Fence inserts speculation barriers into the \gls{BPF} program, which reduce
the instruction-level parallelism inside the \gls{BPF} program. However, the user
and kernel code that calls the \gls{BPF} programs, as well as the kernel helper
functions invoked by the \gls{BPF} programs, are unaffected by this change and
therefore still execute at maximum performance. As most applications only spend
a small part of the CPU time executing \gls{BPF} code, we expect the real-world
overhead to be small, which is also supported by our evaluation. In any case,
Veri\-Fence enables unprivileged users to use \gls{BPF} at all, as their programs
were rejected prior to Veri\-Fence. This saves them from having to implement
their logic in user space, which would then require very expensive user/kernel
switches \cite{gerhorst_anycall_2021}. We therefore deem Veri\-Fence's
negative impact on \gls{BPF}'s performance the more favorable trade-off.

Further, the number of barriers inserted by Veri\-Fence is reduced in
comparison to more naive approaches, as barriers are only inserted when the
verifier actually detects unsafe behavior that could enable a transient
execution attack on the kernel. In addition, verification of a speculative path
is cut short when a barrier is already present. This not only reduces the
verification time but also helps to keep the number of barriers inserted small.
At the same time, this approach avoids complex compile-time analysis, which
would not be practical for \gls{BPF}. Our evaluation supports that this approach
is precise even though it is simple -- we find that Veri\-Fence
inserts a lot fewer barriers than the Spectre-STL defenses already present in
Linux v6.5.

Aside from the execution-time overheads, Veri\-Fence only impacts the
verification time (in comparison to a vulnerable system) for the programs that
would otherwise have been rejected. For these programs, Veri\-Fence
continues to explore the remaining architectural and speculative paths after
discovering unsafe behavior. We do not make any assumptions about the size of
the \gls{CPU}'s speculation window, following the reasoning
from \cite{cauligi_sok_2022}. However, if the hardware were to expose this
information reliably, Veri\-Fence could use it to significantly reduce the
verification time because the verifier indeed knows the exact microarchitecture
on which the code will run. To limit verification time in our prototype, we
selectively cut explored speculative paths short by inserting a barrier
prematurely when we approach the verifier's configurable complexity limits
(e.g., number of instructions simulated, number of branches followed\footnote{For our
benchmarks, we increase the existing limits by a factor of 4 to successfully
verify all real-world \gls{BPF} programs (designed for verification without
Spectre) with full defenses. For Parca, we increase the limit by a factor of
32.}). Using this simple heuristic, we can even verify the largest \gls{BPF}
application programs accepted by the Linux v6.5 verifier without defenses. In
summary, there are multiple approaches that would allow us to reduce the number
of barriers even further without impacting verification complexity. Because
verification is usually not part of the user application's hot
path \cite{craun_enabling_2023} we focus on the execution-time overhead in our
evaluation.

In summary, Veri\-Fence improves the performance of unprivileged user applications
by allowing them to use \gls{BPF} at all. At the same time, our approach has no
impact on privileged users, who can still dynamically disable Veri\-Fence
per-program at execution time if they choose to manually check for gadgets.

\section{Implementation}
\label{sec:org6c8b488}

We implement Veri\-Fence for the Linux \cite{kroah-hartman_linux_2023}
\gls{BPF} verifier and publish our patches\footnote{\anon[Reference omitted for
  double-blind
  review.]{\url{https://sys.cs.fau.de/verifence-linux}}}
% todo: add to i4wp:/proj/.../alias.map
consisting of less than \num{1000} source line changes under an open-source
license. The majority of changes merely restructure the existing \gls{BPF}
verifier to support our design. While we use Linux v6.5, our approach is not
fixed to only apply to this specific Linux version. Instead, it even benefits
from future improvements to the precision of the verifier's architectural
analysis (which also reduce false-positives for Veri\-Fence) and \gls{JIT}
compiler. From a practical perspective, we are not aware of any fundamental
reasons for which the patches could not be merged into upstream Linux.

To make our approach portable, we introduce a distinction between speculation
barriers (verifier-internal \gls{BPF} bytecode instructions) against Spectre-STL,
and \gls{BPF} speculation barriers against Spectre-PHT into the kernel
(\texttt{nospec\_v4} and \texttt{nospec\_v1} respectively). The verifier inserts both barriers
into the bytecode, and the \gls{JIT} compiler backends then either drop or
lower the instructions based on the architecture's configuration. For example,
ARM64 does not require speculation barriers to defend against Spectre-STL due to
its firmware defenses \cite{noauthor_firmware_2018} while x86-64 does require
\texttt{lfence} instructions for both Spectre-STL (unless \acrlong{SSBD}, \acrshort{SSBD}, is
active) and Spectre-PHT. This approach is also compatible with research that
proposes address-specific speculation barriers like \texttt{protect}
from \cite{vassena_automatically_2021}. In summary, vulnerability- and even
address-specific speculation barriers allow the verifier to remain
architecture-agnostic while not impacting performance on architectures that are
not affected.

\section{Evaluation}
\label{sec:orgf874708}
We evaluate Veri\-Fence using both static analysis and real-world execution time
benchmarks. Regarding the former, we analyze whether Veri\-Fence successfully
protects the programs rejected by the upstream Linux v6.5 verifier and how many
speculation barriers it inserts. Regarding the latter, we analyze the
performance impact of the speculation barriers on three popular
performance-critical applications of \gls{BPF}. As discussed in
Section~\ref{sec:apl-to-unpriv-bpf}, our results carry over to future
unprivileged \gls{BPF} programs even if most of the real-world \gls{BPF} programs used today
(and therefore also in our evaluation) still require root privileges.

\subsection{Static Analysis}
\label{sec:org1bd0dd5}
First, we apply Veri\-Fence to all \gls{BPF} programs from our motivating
analysis of the \gls{BPF} program landscape from Section \ref{sec:orgdd58b36}.
Veri\-Fence successfully applies defenses to all real-world application
programs in our dataset and the number of barriers it inserts is insignificant
in comparison to the upstream Spectre-STL defense.

\subsubsection{Program Rejections}
\label{sec:orge1cdc28}

\begin{figure}
\centering
\includegraphics[width=\columnwidth]{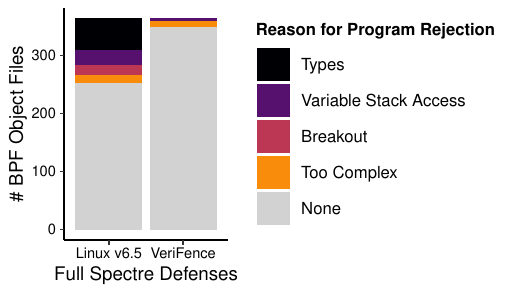}
\caption{Veri\-Fence successfully applies to all real-world application programs
  without modifications. Analyzing all 364 object files,
  defenses only fail to apply to \qty{4}{\percent}
  which is a significant improvement over the upstream verifier's
  \qty{31}{\percent}. Veri\-Fence only rejects
  15 test programs from the Linux
  selftests which exhibit unverifiable architectural behavior that could easily
  be avoided in real applications.\label{fig_verifence_load_problems}}
\end{figure}

Figure~\ref{fig_verifence_load_problems} compares the number and type of verification
errors with Veri\-Fence to the upstream Linux defenses (in both cases the
Spectre-STL defenses are also active). As expected, Veri\-Fence is able to
successfully protect all \gls{BPF} programs from our application group and almost
all programs from the full dataset. The remaining programs are all from the
Linux kernel selftests which Veri\-Fence rejects for one of two reasons:
\begin{itemize}
  \item \textbf{Variable Stack-Access
        (5/364):}
        The verifier currently does not implement defenses against Spectre-PHT
        in the light of variable stack accesses to simplify the
        implementation \cite{matei_01f810_2021}, our modified verifier
        therefore still rejects them. We deem an extension to resolve this out
        of scope as one can easily avoid variable stack accesses in real-world
        \gls{BPF} programs. Our evaluation confirms that rejection due to
        variable stack accesses is not a problem for any of the application's
        \gls{BPF} programs.
  \item \textbf{Too Complex
        (10/364):}
        The Linux kernel selftests contain object files with very large programs
        to test the verifier's complexity limits. Because Veri\-Fence has to
        verify additional speculative program paths, verification can fail if
        the existing program was already designed to be as large as possible. If
        users were to encounter this in the real-world, they could easily
        circumvent it by splitting their program into multiple smaller programs
        that call each other using \gls{BPF} tail calls.
\end{itemize}
In summary, \textbf{VeriFence successfully verifies all \gls{BPF} programs from real
  applications, thereby solving the problem described in Section~\ref{sec:problem}.}
Further, the remaining theoretical reasons for failed verification are easy to
understand and circumvent by \gls{BPF} program developers.

\subsubsection{Speculation Barriers}
\label{sec:org2c53d09}

While it is Veri\-Fence's desired outcome to enable more programs to be
successfully loaded into the kernel, it also has the potential downside of
slowing down those programs because of the speculation barriers it inserts for
defense. In this section and the
following real-world performance evaluation, we analyze the extent to which
these barriers affect performance.

\begin{figure}
\centering
\includegraphics[width=\columnwidth]{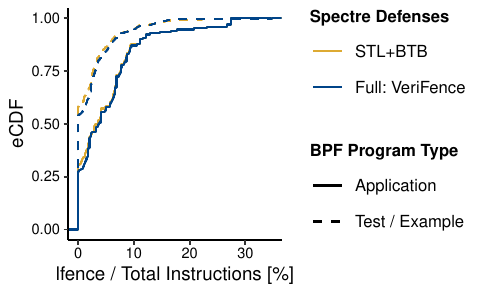}
\caption{Veri\-Fence inserts only insignificantly more speculation barriers than the Spectre-STL defenses present in the upstream Linux kernel.\label{fig_verifence_lfence}}
\end{figure}

First, we repeat the analysis from Section~\ref{sec:org1713803} for Veri\-Fence,
analyzing the number of speculation barriers it inserts in comparison to the
Spectre-STL defenses. Figure~\ref{fig_verifence_lfence} shows the fraction of \gls{BPF}
programs with less than \(X\) percent of speculation barriers for Veri\-Fence
and Spectre-STL defenses (excluding the
15 rejected
programs from the previous section). Veri\-Fence only inserts barriers when
the program would otherwise have been rejected by the verifier, therefore the
overall number of barriers it inserts is small.

In summary, we find that Veri\-Fence has a low expected performance overhead for a
diverse set of applications. In the following section, we will back up this
claim using performance benchmarks for three real-world applications using
\gls{BPF}.

\subsection{Real-World Performance Evaluation}
\label{sec:orgcb20806}

In this section, we analyze the real-world application overhead of
Veri\-Fence for over 50 distinct \gls{BPF} programs. We focus on event
tracing, continuous profiling, and packet processing, as these are three of the
most popular performance-critical applications of \gls{BPF}. As more complex
programs are more likely to exhibit unsafe transient behavior, we select the
Parca Continuous
Profiler \cite{javier_honduvilla_coto_parca-devparca-agent_2023} and the Loxilb
Network Load Balancer \cite{packetcrunch_github_user_loxilb_2024} to analyze
Veri\-Fence's overhead in extreme cases.

We run all our execution-time benchmarks in a Debian 11 GNU/\allowbreak{}Linux
system with a v6.5.11 kernel on a 6-core \qty{2.8}{\giga\Hz} Intel CPU from 2017
(Intel Core i5-8400). We disable dynamic voltage and frequency
  scaling~(DVFS) as the execution-time measurements would otherwise be highly
  dependent on the current system load~\cite{wysocki_cpufreq_2024}, which
  is not relevant to our evaluation and therefore only hinders reproducibility.
In our graphs, we use standard boxplots \cite{wickham_ggplot2_2024} showing the
first, second (i.e., median), and third quartile.

\subsubsection{Event Tracing}
\label{sec:org4a1e252}

Tracing is one of the most popular applications of \gls{BPF} useable both for
performance debugging and continuous monitoring in production. It allows users
to gain valuable insights into kernel execution with little to no impact on
production workloads. Still, their use poses security risks as, without Spectre
defenses, \gls{BPF} programs can easily introduce arbitrary-read gadgets into the
kernel by mistake, therefore jeopardizing the security of the whole system. In
this section, we analyze the execution-time overhead Veri\-Fence has for
BCC's \texttt{libbpf}-based tracers when monitoring a system that runs
Memcached \cite{dormando_memcached_2020} together with the
\texttt{memtier\_benchmark} \cite{gottlieb_memtierbenchmark_2022} load generator.
Client and server each use three threads and communicate using Memcached's
binary protocol. We perform \num{15000} requests taking
\qty{18}{\second} on
average and repeat each test
100~times.

\begin{figure*}
\centering
\includegraphics[width=\linewidth]{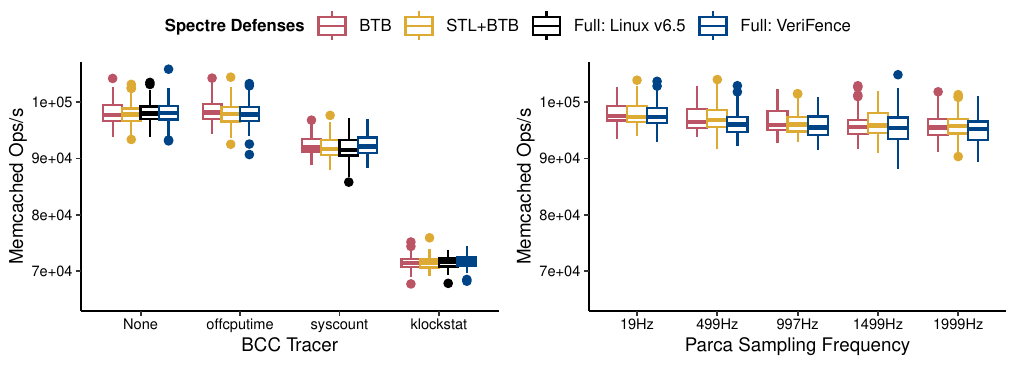}
\caption{Veri\-Fence does not affect Memcached performance when using any of the
  43
  analyzed BCC tracers (of which the three most CPU-intensive are displayed in
  the left plot) or the Parca Continuous Profiler (right plot) to monitor the
  workload.\label{fig_tracer}}
\end{figure*}

We analyze
43 BCC
tracers\footnote{This includes the 39 \gls{BPF} objects from Section~\ref{sec:orgdd58b36}
but the number is higher here because some \gls{BPF} objects are not designed to
be loaded using \texttt{bpftool}.}. Out of these tracers,
21 require
Veri\-Fence to be successfully used with Spectre-PHT defenses. When running
the tracers alongside the workload, we find that
22 of
the tracers record at least 10 events per second and
9 at
least 1000. We conclude that most of the tracers are invoked frequently and
collect a reasonable amount of information in this setup. However, the CPU time
the tracers spend executing the \gls{BPF} programs themselves is small. Out of the
analyzed tracers,
9 tracers
spend more than \qty{0.1}{\percent}, and
3 tracers
spend more than \qty{1}{\percent} of CPU time executing \gls{BPF} code.

The left plot of Figure~\ref{fig_tracer} shows the impact that the three most
CPU-intensive tracers have on Memcached's performance in four different system
configurations. Without Veri\-Fence, the verifier cannot successfully apply
Spectre defenses to \texttt{offcputime}'s \gls{BPF} program, thus the baseline system
configuration (\emph{Full: Linux v6.5}) is missing for this tracer. Defenses do
successfully apply to \texttt{syscount} and \texttt{klockstat} without Veri\-Fence, but we still
measure the overhead with Veri\-Fence for completeness. Overall, Veri\-Fence does
not have any measurable impact. We also analyze the CPU time spent executing
\gls{BPF} code and find no measurable difference between Veri\-Fence and the other
configurations even when we run each test for a total duration of
\qty{30}{\minute}.
This is expected as tracers are designed to not impact production workloads by
consuming as few resources as possible. They can therefore benefit from the
security benefits Veri\-Fence offers without experiencing increased overheads.

\subsubsection{Continuous Profiling}
\label{sec:org9fe1dca}

The second real-world \gls{BPF} application we analyze is continuous profiling.
Here, a \gls{BPF} program that records the current user and kernel stack trace
is invoked by a timer interrupt at a configurable frequency. The data can later
be analyzed to find performance bugs, for example using flame graphs \cite{gregg_flame_2016}. Traces are either collected continuously in production
(sampling frequencies below \qty{150}{\Hz}) or on-demand by developers (usually
\qtyrange{500}{2000}{\Hz}).

For this benchmark, we run the Parca Continuous Profiler alongside the Memcached
workload from the previous section. The results are displayed in
Figure~\ref{fig_tracer} on the right. As the \gls{BPF} program that Parca uses to
collect the stack samples is relatively complex, it cannot be successfully
defended against Spectre-PHT without Veri\-Fence. Therefore, the baseline
(\emph{Full: Linux v6.5}) is missing from the figure. Analyzing the amount of CPU
time spent in \gls{BPF} code for this benchmark, we find that it is only
\qty{1.0}{\percent} even
when the maximum sampling frequency we deem reasonable is used. The main
overhead of the Parca tracer therefore originates from the code that
post-processes the samples in user space. From this, it follows that both the
Spectre-STL defenses and the Spectre-PHT defenses using Veri\-Fence do not
impact Memcached's performance even though they increase the execution time of
the \gls{BPF} programs by
\qty{16}{\percent} and
\qty{62}{\percent} respectively.
In comparison to a vulnerable system, Veri\-Fence only reduces Memcached's
throughput by
\qty{0.8}{\percent} at
\qty{1999}{\Hz}. At frequencies below \qty{150}{\Hz}, the overhead is no longer
measurable
(\qty{0.1}{\percent} for
\qty{19}{\Hz} in this particular run). In summary, Veri\-Fence can be
applied to Parca's \gls{BPF} program without overheads to the workload.

\subsubsection{Network Load Balancing}
\label{sec:org708cfd2}

High-performance packet processing is the application \gls{BPF} was originally
developed for and is still one of the most popular use cases \cite{yang_heels_2023,dasineni_open-sourcing_2018,vieira_fast_2020,hoiland-jorgensen_express_2018,crampton_introducing_2020,graf_cni_2021,packetcrunch_github_user_loxilb_2024}.

In this benchmark, we replicate a real-world scenario where multiple Docker
containers communicate with each other through the \gls{BPF}-based Loxilb network
load balancer \cite{packetcrunch_github_user_loxilb_2024}. To stress the load
balancer, we extend the benchmarks the Loxilb upstream project includes. To
achieve the maximum possible CPU utilization, we do not run the containers on
different physical machines. Therefore, the overhead we measure represents the
upper bound as the performance is not limited by the speed of the networking
interface but only by the CPU's speed.

\begin{figure*}
\centering
\includegraphics[width=\linewidth]{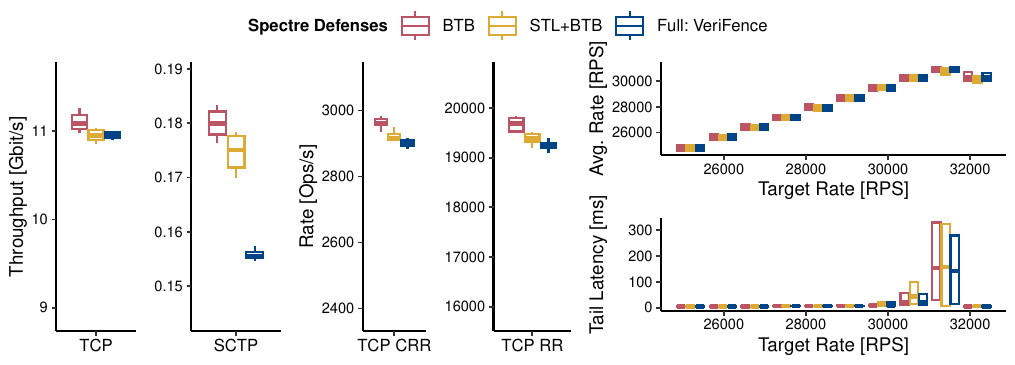}
\caption{\gls{TCP} and \gls{SCTP} throughput, TCP \acrlong{RR}- (\acrshort{RR}) and \gls{CRR} rate, and \texttt{nginx} HTTP tail latency achievable over a \gls{BPF}-based Loxilb load balancer. The upstream Linux v6.5 verifier cannot successfully apply Spectre defenses to Loxilb's \gls{BPF} program, therefore the baseline (\textit{Full: Linux v6.5}) is missing.\label{fig_loxilb}}
\end{figure*}

The four left plots of Figure~\ref{fig_loxilb} show the maximum throughput and request
rate for \gls{TCP} and \gls{SCTP}. We normalize each scale to display between
\qty{80}{\percent} and \qty{105}{\percent} of the respective performance
achievable with only the Spectre-BTB defenses active (the default in Linux v6.5). For
\texttt{netperf} \cite{wenzel_netperf_2020} and \texttt{iperf3} \cite{mah_iperf3_2020}, we
use one client thread but confirm that multiple threads do not impact our
conclusions in separate experiments (e.g., using \texttt{iperf} v2.0.14a). We run each
benchmark for
\qty{2}{\minute} and
repeat the measurement
10 times.

We observe that Veri\-Fence only has a small impact on TCP throughput and
average latency. Compared to a vulnerable system, both Veri\-Fence and
\emph{STL+BTB} defenses reduce the TCP throughput by
\qty{1.1}{\percent} to
\qty{1.2}{\percent}.
As expected, the \gls{RR} rate is affected the most by Veri\-Fence with
\qty{2.1}{\percent}
overhead while the \gls{CRR} rate only decreases by
\qty{1.8}{\percent}.
In summary, the main overhead stems from the Spectre-STL defenses, while Veri\-Fence only has a negligible impact.

\gls{SCTP} is a message-based protocol that still ensures reliable
transportation. In comparison to \gls{TCP} and \gls{UDP}, it offers native support
for multihoming. However, because it is not as popular, the Linux kernel does
not offer a \gls{BPF} helper function to recompute the checksum for packets that
are redirected. Therefore, Loxilb has to recompute the \gls{SCTP} checksum in
\gls{BPF} itself. Packet redirection thus consumes much more CPU time and the
throughput is reduced by over \(60 \times\) in comparison to TCP. This also
causes increased relative overheads when we activate the Spectre-STL defenses
(\qty{2.7}{\percent})
and Veri\-Fence
(\qty{14}{\percent}).
To improve \gls{SCTP}'s performance, it would be the most promising to either
disable checksum calculation \cite{nik-netlox_github_user_sctp_2023}, or create
a kernel helper function. Compared to the potential speedup by avoiding checksum
recalculation in \gls{BPF} altogether, the speedup achieved by disabling
Veri\-Fence is insignificant for~\gls{SCTP}.

In summary, Veri\-Fence does not bottleneck the \gls{TCP} and \gls{SCTP}
performance achievable over the load balancer. This is the case especially
because the overhead would be even lower if the containers were not colocated
on the same machine.

For our final benchmark, we run two \texttt{nginx} servers serving a
% do not use  \qty{1}{\kibi\byte} as acm taps does not support it
1\,KiB payload and connect to them through a single Loxilb load balancer instance. We
use the \texttt{wrk2} load generator \cite{rova_wrk2_2023} with two threads to target
a specific HTTP request rate and analyze the resulting tail latency (99nth
percentile) and average rate. We use \texttt{wrk2} because it does not suffer from
coordinated omissions. With two servers, the maximum achievable rate is approx.
\qty{31}{\kilo RPS}. Based on this, we scale the target load from
\qty{80}{\percent} to \qty{105}{\percent} in increments of \qty{0.1}{\percent}
and then group the data into bins of \qty{2.5}{\percent}. We run each test for
\qty{1}{\minute} and
repeat the measurement
10 times,
therefore each bin contains \num{250} data points. We omit outliers from this
plot.

The right plots of Figure~\ref{fig_loxilb} show the achieved average rate and tail
latency for different target rates. As expected, we observe increased variation
for all configurations when running the system close to \qty{31}{\kilo RPS} as
small variations are easily amplified due to system-overload. Importantly,
Veri\-Fence does not cause increased tail latencies. This is because Veri\-Fence
merely increases the time needed to process each request by a small amount, but
does not cause any unpredictable spikes in the processing time.

\subsubsection{Summary}
\label{sec:orgc67c918}

\textbf{Veri\-Fence does not affect most of the applications we analyze in a
measurable way.} Event tracers appear unaffected, which is particularly
important as developers often write small tracing programs
ad-hoc \cite{xu_bpftrace_2024} and therefore are unlikely to exhaustively
review their programs for gadgets. For more complex \gls{BPF} programs as used by
the Parca Continuous Profiler and the Loxilb Network Load Balancer, we find that
both the Spectre-STL defenses and Veri\-Fence can affect the CPU time spent
executing \gls{BPF} code, however, \textbf{this usually does not impact application
performance, showing that it should only be of secondary concern.}
Instead, it is much more important to be able to use \gls{BPF} in the first place. In any case, users can
always disable Veri\-Fence per--\acrshort{BPF}-program at runtime to trade
security for performance.

\section{Related Work}
\label{sec:orgd696d15}

In this section, we discuss alternatives to \gls{BPF} and then relevant
alternative techniques to defend against transient execution attacks.

\subsection{High-Performance \protect\acrshort{IO}}
\label{sec:orgba764a4}

\gls{BPF} implements safe kernel
extensions \cite{fahndrich_language_2006,hunt_singularity_2007,zhou_userspace_2023}
for Linux. While it has numerous applications outside of high-performance \gls{IO}
\cite{lian_ebpf-based_2022,kim_triaging_2022,jia_programmable_2023,yang_redis_2022,heo_patchset_2023},
asynchronous \gls{IO} and kernel-bypass can replace it in some cases.

\subsubsection{Asynchronous \protect\acrshort{IO}}
\label{sec:org5632e63}

To improve performance over traditional system calls, asynchronous \gls{IO}
\cite{alexander_van_der_grinten_managarm_2024} is implemented by
\texttt{aio} \cite{bhattacharya_asynchronous_2003} and more recently by
\texttt{io\_uring} \cite{axboe_io_uring_2019} in
Linux \cite{didona_understanding_2022}. Being closely related to system-call
batching \cite{zadok_efficient_2005,gerhorst_anycall_2021}, it amortizes mode
switches over multiple operations. However, both do not improve upon the latency
of traditional system-calls because data still has to pass through the \gls{OS}
network or storage stack to reach the application. In contrast, \gls{BPF} allows
applications to process (or discard) data directly on the CPU where it is first
retrieved \cite{zhong_xrp_2022}.

\subsubsection{Kernel-Bypass}
\label{sec:org860ad3f}

The most prominent implementations of ker-\allowbreak{}nel--bypass are
\acrshort{DPDK} \cite{noauthor_dpdk_2024,rosen_network_2017} and
SPDK \cite{noauthor_spdk_2024,yang_spdk_2017}. By giving user applications
direct access to the hardware, kernel-bypass can reduce both \gls{IO} latency and
throughput. However, in contrast to \gls{BPF}, kernel-bypass does not integrate
cooperatively with the existing networking stack and requires dedicating full
\gls{CPU} cores to busy-looping for low-latency packet processing (hurting
power-proportionality). Further, when kernel-bypass and regular applications are
colocated on a server, kernel-bypass reduces performance for regular applications
because packets have to be re-injected into the kernel networking stack to reach
them \cite{hoiland-jorgensen_express_2018}. For this reason, Meta reportedly
uses a \gls{BPF}-based load balancer (similar to Loxilb from our evaluation)
instead of kernel-bypass \cite{dasineni_open-sourcing_2018}.

\subsection{Transient-Execution--Attack Defenses}
\label{sec:orgf1b72f1}

To defend against transient execution attacks, Veri\-Fence relies on
compiler-based defenses. Alternative approaches partition resources
(OS-based) or attempt to implement side-channel--resistant
transient execution that is still low-overhead (hardware-based).

\subsubsection{Compiler-based Defenses}
\label{sec:org18f460f}

To the best of our knowledge, Linux's \gls{BPF} verifier is the only widely
deployed sandbox that is fully hardened against the known Spectre
vulnerabilities without relying on process isolation. While incomplete Spectre
defenses have been implemented for some sandboxes (e.g.,~index masking and
pointer poisoning for JavaScript~\cite{pizlo_webkit_2018}), all
production Java~\cite{naseredini_systematic_2021},
{\gls{Wasm}}~\cite{reis_site_2019,wasmtime_security_2024}, and
JavaScript~\cite{bynens_untrusted_2018,titzer_year_2019} runtimes still
rely on process isolation for full mitigation, as recommended by
Intel~\cite{noauthor_managed_2018}. However, the kernel cannot apply
process isolation to \gls{BPF} without reducing its performance
significantly \cite{gerhorst_anycall_2021}. The Windows \gls{BPF} verifier
(Prevail \cite{jowett_vbpfebpf-verifier_2021}) does not implement Spectre
defenses as of April 2024 \cite{gershuni_use_2022}.

There exist theoretical works that can potentially detect Spectre gadgets more
precisely than the \gls{BPF} verifier (even with our
extension) \cite{cauligi_sok_2022,patrignani_exorcising_2021}. However, these
cannot be readily applied to \gls{BPF} because they make simplifying assumptions
in their
implementation \cite{guarnieri_spectector_2020} or are
too complex to
implement \cite{vassena_automatically_2021}
(and would,
therefore, further increase the risk of security-critical bugs). The common
C/C++ toolchains do not offer reliable, architecture-agnostic Spectre defenses
that are transparent to the
users \cite{noauthor_managed_2018,noauthor_visual_2021,kocher_spectre_2018,beyls_d49073_2018,beyls_d49070_2018,beyls_d41760_2018,jeanpierre_mitigating_2020},
therefore the \gls{BPF} verifier must apply defenses to the programs. Most
related works on \gls{BPF} focus on architectural security but do not consider
transient execution
attacks \cite{mahadevan_prsafe_2021,nelson_proof-carrying_2021,nelson_exoverifier_2022,lu_moat_2023}.
In future work, the number of barriers Veri\-Fence inserts could be further
reduced by using \gls{ALU} instructions to detect misprediction and erase
sensitive data (e.g., using \acrlong{SLH},
\acrshort{SLH} \cite{carruth_llvm-dev_2018,zhang_breaking_2022,shivakumar_spectre_2023,carruth_speculative_2024,patrignani_exorcising_2021}
and similar
techniques \cite{oleksenko_you_2018,ojogbo_secure_2020,borkmann_bpf_2019}).

While browsers to date rely on process isolation to prevent \gls{Wasm} programs
from exploiting Spectre \cite{reis_site_2019,ojogbo_secure_2020}, there is
some work on compiler-based defenses in specific runtimes. However, we find that
they are either incomplete (i.e., Wasmtime \cite{wasmtime_security_2024}, V8
\cite{bynens_untrusted_2018,mcilroy_spectre_2019}) or specific to
\gls{Wasm}/user space (i.e., Swivel \cite{narayan_swivel_2021}) and do not apply
to \gls{BPF} in the kernel.

\subsubsection{\protect\acrshort{OS}-based Defenses}
\label{sec:orgec827f6}

To defend against transient execution attacks, most works rely on coarse-grained
partitioning of resources \cite{hertogh_quarantine_2023,hofmeier_dynamic_2022}
with support from the \gls{OS} (e.g., address-space
isolation \cite{behrens_efficiently_2020} and core
scheduling \cite{noauthor_core_2023}). However, address-space isolation cannot
be applied to \gls{BPF} without limiting its performance even
further \cite{gerhorst_anycall_2021}. Applying \glspl{MPK} to \gls{BPF} against
Spectre appears to be a promising direction for future
research \cite{noauthor_refined_2022,noauthor_managed_2018}. Existing work on
the topic does not take transient execution attacks into
account \cite{lu_moat_2023} but assumes the \gls{BPF} verifier (e.g., our work)
implements defenses.

\subsubsection{Hardware-based Defenses}
\label{sec:org88298fc}

As of April 2024, there exists no practical high-performance processor
implementation that is not vulnerable to
Spectre \cite{amd:software:2023,intel:affected:2024,arm:speculative:2024}.
Instead, vendors continue to recommend compiler-based defenses against all
cross- and in-domain transient execution
attacks \cite{noauthor_refined_2022,noauthor_optimized_2020}. To date,
complete protection from microarchitectural timing side-channels is not possible
without significant changes to the \gls{ISA} and processor
design \cite{yu_stt_2019,xiong_survery_2021,heiser_towards_2020,buckley_proving_2023}.

\section{Conclusion}
\label{sec:org27deb8b}

This work has presented Veri\-Fence for Linux \gls{BPF}, a practical and
easy-to-apply enhancement to the only software-based sandbox resilient against
Spectre. Veri\-Fence is \emph{sound}, \emph{precise}, and \emph{lightweight}.
First, it \emph{soundly} combines speculation barriers and static analysis in a
way that does not increase the kernel's attack surface. Second, it
\emph{precisely} prevents unsafe transient behavior thereby reducing the number
of rejected programs from
\qty{31}{\percent}
to
\qty{4}{\percent}
(with only test programs from the Linux selftests remaining). Finally, as our
real-world performance evaluation demonstrates, it is \emph{lightweight} because
it does not impact \gls{BPF} invocation latency, which is of particular
importance to \gls{BPF} as it complements user space by offering minimum-overhead
transitions.

\begin{acks}
  We thank Daniel Borkmann, Alexei Starovoitov, Benedict Schlüter, Marco
  Guarnieri, Georg Julius Liefke, and Anil Kurmus for taking the time to review
  our kernel patches and giving their thoughtful feedback on the project. We are
  grateful to the anonymous reviewers and our shepherd for their insightful
  comments. This work is supported by the German Research Foundation (DFG) --
  project numbers 539710462 (DOSS), 502615015 (ResPECT), 502228341 (Memento),
  and 465958100 (NEON).
\end{acks}

\bibliographystyle{ACM-Reference-Format}
\bibliography{references}

\end{document}